\documentclass[aip,12pt,reprint]{revtex4-1}

\usepackage{graphicx}

\begin{document}

\title{Experimental observation of induced stochastic transitions in multi-well potential of RF SQUID loop} %
\author{O.G. Turutanov}
\email{turutanov@ilt.kharkov.ua} %
\affiliation{B. Verkin Institute for Low Temperature Physics and
Engineering, National Academy of Sciences of Ukraine, 47 Lenin ave.,
Kharkov 61103, Ukraine}%
\author{V.Yu. Lyakhno}
\affiliation{B. Verkin Institute for Low Temperature Physics and
Engineering, National Academy of Sciences of Ukraine, 47 Lenin ave.,
Kharkov 61103, Ukraine}%
\author{V.I. Shnyrkov}
\affiliation{B. Verkin Institute for Low Temperature Physics and
Engineering, National Academy of Sciences of Ukraine, 47 Lenin ave.,
Kharkov 61103, Ukraine}%

\begin{abstract}

An amplification of a weak low-frequency harmonic signal is
experimentally observed in a single-junction-superconducting quantum
interferometer (RF SQUID loop) due to stochastic transitions between
two or more metastable states of the loop under the applied noise
flux of varying intensity (the effect of stochastic resonance, SR).
In addition to the usual scenario of the SR in a bistable system
with Gaussian noise, the transitions between multiple metastable
states of the multi-well SQUID loop potential under the influence of
a binary noise is observed. This can be interpreted as a kind of
noise "spectroscopy" of the metastable states of the SQUID loop with
different values of the trapped magnetic flux.
\end{abstract}



\maketitle

\section{Introduction}

      The Superconducting Quantum Interference Devices (SQUIDs) are
the base element of the most currently sensitive magnetometers
widely used in both physical experiments and applications (medicine,
geophysics, industry). The sensitivity of the SQUIDs and their
quantum analogues, SQUBIDs (Superconducting QUBIt Detectors), has
practically reached the quantum limitation \cite{1,2,3}. However,
with increase of the temperature  $T$  and the quantizing loop
inductance $L$, the uncertainty of the noise flux rises thus
resulting in a degradation of the energy resolution (sensitivity).
This issue is the most severe for high-$T_c$ SQUIDs cooled down to
the liquid nitrogen temperature level where the quantizing loop
inductance and, correspondingly, the sensitivity to the magnetic
field have to be considerably reduced. However, as shown in papers
\cite{4,5,6,7,8,9}, due to the same thermodynamic fluctuations and
the external noise, the sensitivity of the magnetometers can be
essentially enhanced by using the stochastic resonance (SR) effect.

      The SR phenomenon discovered in the early 1980s \cite{10,11}
manifests itself in a non-monotonic rise of the response of a
non-linear, often bi-stable, system to a weak periodic signal whose
characteristics (amplitude, signal-to-noise ratio) become better at
the system output. Actually, to enable the SR in a specific system,
it is enough that the time duration of the system being in one of
its metastable states (MS) (the residence time) would be a function
of the noise intensity. The SR effect has been found in numerous
natural and artificial systems, both classic and quantum; the
analytical approaches, the criteria and quantifiers to estimate the
noise-induced ordering were elaborated \cite{12,13}. For the
aperiodic systems with strong dissipation (which are mostly explored
theoretically and experimentally), the "stochastic filtration" (SF)
is rather more appropriate term than the widely accepted "stochastic
resonance" \cite{14}. Practically all the high-$T_c$ SQUIDs belong
here.

      Although a noticeable number of the theoretical and modeling
studies of the SR in the superconducting loop are published, there
is still a lack of the experimental investigations of the stochastic
dynamics in SQUIDs \cite{15} (e.g., \cite{4,5,6}). Therefore some
questions still remain in this field, including possible practical
applications. For example, is it possible to enhance substantially
the sensitivity of the high-$T_c$ SQUIDs with SR by an optimal
choice of their parameters in the area of strong fluctuations  $L\ge
L_{F} =\Phi _{0}^{2} /(4\pi^{2}k_{B}T)$? Here $L_{F} $  is the
fluctuation inductance,  $L_{F} \sim 10^{-10}$  H at  $T=77$  K,
$\Phi_{0}=2.07\cdot 10^{-15}$  Wb is the superconducting magnetic
flux quantum,  $k_{B}$  is the Boltzmann constant. Additionally, it
should be noted that even if the system has many states, usually the
transitions between only two adjacent MSs are considered in the
context of the SR and SF. The noise type and statistics may
introduce their specifics in the stochastic dynamics of the
interferometers.

      In this work we experimentally observed an amplification of a
weak harmonic low-frequency signal in an RF SQUID loop due to
stochastic transitions between two and more metastable states of the
loop induced by the Gaussian or binary noise of varying intensity
(the stochastic filtration effect).

\section{RF SQUID dynamics and experimental}

      In the absence of fluctuations, the number of the local minima
of the quantizing loop potential energy
$u(x,x_{e})=(x-x_{e})^{2}/2-\frac{\beta_{L}}{4\pi^{2}}\cos(2\pi x)$
(the number of the loop MSs) is defined by the dimensionless
parameter of non-linearity $\beta_{L} =2\pi LI_{c}/\Phi_{0}$:
$n\approx \, 2\beta_{L}/\pi$. Here  $x_{i} =\Phi_{i}/\Phi_{0}$ and
$x_{i} =\Phi_{i}/\Phi_{0}$  are the dimensionless internal and
external fluxes, correspondingly, the energy is normalized to
$\Phi_{0}^{2}/2L$, while $I_c$ is the contact critical current.

      In our experiments, we tested the interferometers with large
$\beta_{L}\approx 7-10$  and the low-impedance ($R\sim 1$ $\Omega$)
Josephson junctions of ScS
(superconductor-constriction-superconductor) type having low
intrinsic capacity ($C\approx 3\cdot 10^{-15}$  F). Such a set of
the parameters determine the overdamped regime which is also typical
for most of the high-$T_c$ SQUIDs.

      Taking into account the smallness of  $C$  and  $R$, we may
neglect the second derivative in the flux motion equation \cite{16}
and reduce it to the form convenient for computations \cite{7,8,9}:

$\displaystyle \frac{dx}{dt} =\frac{1}{\tau _{L} } \left[x_{e}
(t)-x+\frac{\beta _{L} }{2\pi } \sin (2\pi x)\right]$,

\noindent
where  $\tau_{L} =L/R$  is the loop flux decay time. The
equation describes an aperiodic system. The external flux $x_{e}$ is
the sum of the fixed bias flux ($x_{dc}=0.5$) to symmetrize the
potential, the weak low-frequency information signal $x_{s}=a\sin
2\pi f_{s} t$ ($a<<1$) and uncorrelated (white) Gaussian noise
$x_{N} =\xi(t)$, $\left\langle \xi(t)\cdot \xi (t')\right\rangle
=2D\delta (t-t')$, where  $D$  is the noise intensity. However, both
in the calculations and the experiments, the noise is frequency-band
limited by a cut-off frequency. To consider it practically
uncorrelated in the context of the discussed SR model, the cut-off
frequency  $f_{c}$ should sufficiently exceed the signal frequency:
$f_{c}>>f_{s}$. In our experiments, we chose  $f_{s} =37$ Hz and
$f_{c} =50$  kHz. The used instrument generated the Gaussian or the
binary (telegraph) noise originated from a real physical source
(diode).

\begin{figure}[t!]
\centering %
\includegraphics[width = 1.0\columnwidth]{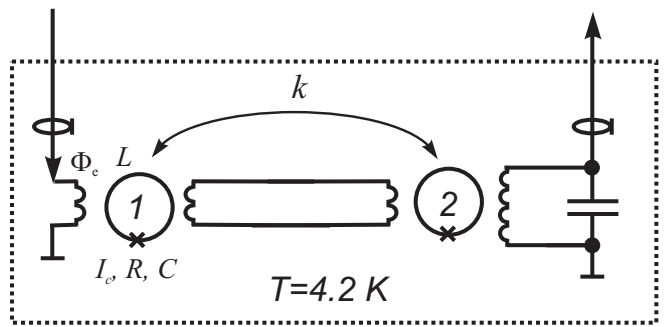}
\caption{\label{fig01} %
The simplified diagram of the measurements. External magnetic flux
$\Phi_{e}$  is applied by a coil to the loop of the interferometer
under test (denoted by number 1). The in-loop flux is measured with
the RF SQUID magnetometer (denoted by number 2) via the
superconducting magnetic flux transformer. The coupling coefficient
between the loops is denoted by  $k$ . The dotted line indicates the
superconducting led shield.
}%
\end{figure}

      The interferometer under test (denoted by 1 in Fig.\,\ref{fig01}) was
designed as a niobium 3D self-shielded toroidal construction with
the adjusted point contact. It coupled to an instrumental RF SQUID
magnetometer (denoted by 2 in Fig.\,\ref{fig01}) via the
superconducting magnetic flux transformer with the interferometer
loop-to-loop flux coupling coefficient  $k=0.05$. The interferometer
design is described in detail in the paper\cite{2}. The spectral
density of the magnetic flux noise (the sensitivity) of the
magnetometer was $S_{\Phi }^{1/2} \approx 2\cdot 10^{-4} \;
\Phi_{0}/Hz^{1/2}$ within the operation frequency band of 2 to 200
Hz. The coupling coefficients, the fluxes and the coil RF currents
were determined from the measurements of the amplitude-frequency and
the amplitude-flux characteristics of the interferometer under test
while changing the in-loop flux within $\pm 5\Phi_{0}$. The
experimental setup is similar ideologically to that reported in
\cite{5} and will be described elsewhere. The measurements were
taken at temperature 4.2 K inside a superconducting shield. The
cryostat, in its turn, was placed into the three-layer mu-metal
shield. The output signal was fed to the spectrum analyzer
Bruel\&Kjer 2033. The number of the instrumentally-averaged spectra
was 16. The readings were taken doubly, with and without the
information signal. The difference between the two spectra obtained
was considered as the result.

\section{Results and discussion}

      The numerical calculations \cite{4,5,7,8,9} show that the spectrum density
of the internal flux in the SQUID loop at the useful signal
frequency rapidly rises, peaks and then slowly decreases with the
increase of the Gaussian noise intensity  $D$, in accordance with
the theory \cite{12,13}.

\begin{figure}[t!]
\centering %
\includegraphics[width = 1.0\columnwidth]{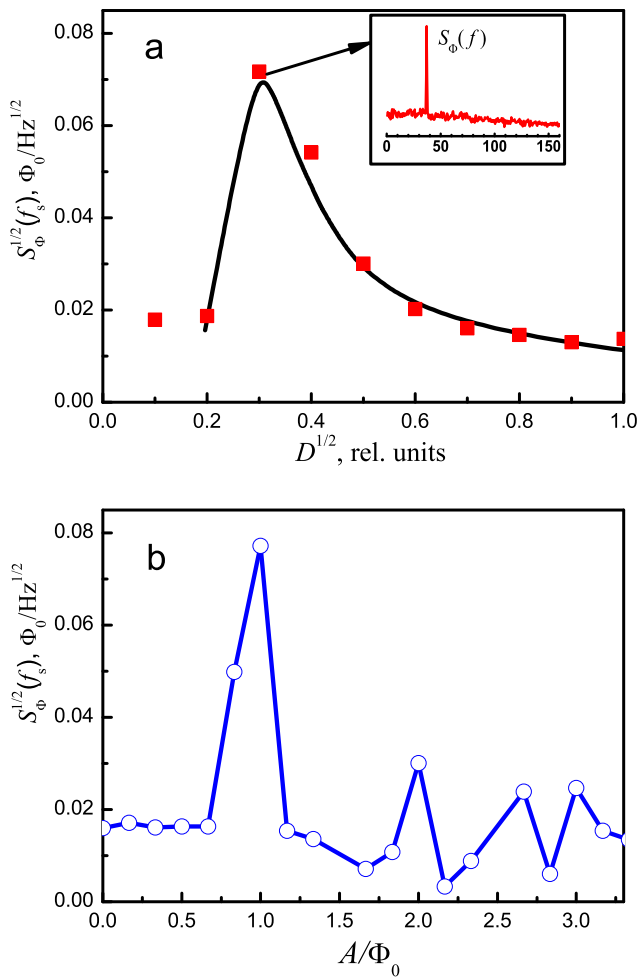}
\caption{\label{fig02} %
(Color online) The amplitude spectral density $S_{\Phi}^{1/2}$  of
the magnetic flux inside the RF SQUID loop at frequency  $f_{s}$ of
the weak harmonic signal as a function of the mean-square amplitude
$D^{1/2}$  of the Gaussian noise (solid line is the numerical
calculation, solid squares are the experimental points) (a) and of
the reduced amplitude  $A/\Phi_{0}$  of the binary noise (hollow
circles are the experimental points) (b). The inset in (a) shows the
experimental spectrum of the output signal (measured in-loop flux)
at the maximum signal gain.
}%
\end{figure}

      Fig.\,\ref{fig02}(a) exhibits the amplitude spectral density of
the flux inside the interferometer loop at the information signal
frequency $S_{\Phi}^{1/2} (f_{s})$  calculated numerically as a
function of the mean-square amplitude of the Gaussian noise
$D^{1/2}$ (solid line) along with the experimental points (solid
squares). The amplitude of the harmonic signal inside the
interferometer under test was $a=0.015$  in  $\Phi_{0}$  units. The
interferometer behavior is typical for the scenario of the SR (more
accurate, SF) in a bi-stable system. The inset in Fig 2(a) displays
the spectrum of the in-loop magnetic flux corresponding to the
maximum signal gain. The average frequency of the noise-induced
transitions between the MSs (Kramers rate \cite{17}) $r_{K} \approx
2f_{s}$  in this point, while the in-loop flux spectrum takes  $1/f$
form. For the chosen, rather large, $\beta_{L}$ , the maximum gain
of about 10 dB was obtained in this experiment.

      The system demonstrates different behavior with the binary
(telegraph) noise instead, whose amplitude is fixed while the phase
is random. The time of the system residence in one of its MSs in the
case of the Gaussian noise changes gradually with the noise
intensity. Under the binary noise, the system remains in a certain
MS during the noise intensity rise until the noise amplitude becomes
sufficient to cause the induced transition to another MS. The
"stochastic synchronization" in this situation occurs between the
useful signal phase and the random phase of the binary noise.

      As it was mentioned above, the SR phenomenon is considered
assuming the two MSs separated by the potential barrier with a small
height to achieve larger gain. For the low-temperature SQUID, this
means that  $\beta_{L}$  should range from 1 to 1.5. However, the
high-$T_c$ SQUIDs cooled down to the liquid nitrogen temperature
level, in which the use of SR could be advantageous, keep
quasi-nonhysteretic behavior up to  $\beta_{L} \approx 3.5$
\cite{18,19,20,21} due to the large variance of the magnetic flux
fluctuations. In this area of strong fluctuations, the number of MSs
can be measured by means of the SR.

      We tested the interferometer with a few MSs for the SR effect
varying the amplitude of the external binary noise.
Fig.\,\ref{fig02}(b) demonstrates the amplitude spectral density
$S_{\Phi}^{1/2}(f_s)$  of the in-loop magnetic flux at the useful
signal frequency  $f_{s}$ as a function of the amplitude  $A$  of
the in-loop binary noise expressed in  $\Phi_{0}$  units. The signal
parameters are the same as above: the frequency  $f_{s}=37$  Hz, the
amplitude of the in-loop flux variation  $a=0.015\; \; (\Phi_{0})$.
The amplitude of the in-loop magnetic noise flux ranged from 0 to
$\sim 3.5\; \Phi_{0}$. It is seen from the plot in
Fig.\,\ref{fig02}(b) that the additional signal gain maxima
corresponding to the stochastic transitions between several MSs of
the loop with the different captured flux are observed while
increasing the noise amplitude. The well-determined amplitude of the
telegraph noise allowed us to perform a kind of "spectroscopy" of
the MSs in the superconducting interferometer loop in the limit of
strong fluctuations. The  $\beta _{L} $  for this interferometer was
estimated by the amplitude-frequency characteristics without an
external noise as about 10.

\section{Summary}

      The SR (or SF, for an overdamped aperiodic system) effect
enables certain gain of a weak periodic signal due to "stochastic
synchronization" of the noise-induced transitions between two or
more metastable states with this useful signal. This scenario is
observed in the experiment with the superconducting quantum
interferometer. To obtain a large enough gain (like, e.g. in
\cite{5}) predicted theoretically (see Refs. in \cite{4,5,9}) and
numerical calculations \cite{8,9}, it is necessary to optimize the
SQUID parameters, particularly, use the interferometer with low
$\beta_{L} \ge 1$ . For the high-$T_c$ RF SQUIDs in the area of
strong fluctuations, the effective double-well potential and the
best stochastic gain of a weak information signal should be observed
at $\beta_{L} \approx 4$ .

      The stochastic gain of a weak periodic signal is observed due
to the noise-induced transitions not only between the two adjacent
metastable current states of the superconducting interferometer loop
but also between several more distant ones. The picture is more
distinct in the case of the binary noise that gives grounds to
consider the procedure as a specific "noise spectroscopy" of the
system metastable states.

\end{document}